\documentclass[aip,apl,reprint]{revtex4-1}
\usepackage{amsmath}
\usepackage{amsfonts}
\usepackage{amssymb}
\usepackage{graphics}
\usepackage{float}
\usepackage{graphicx}
\usepackage{epstopdf}
\usepackage[compact]{titlesec}
\usepackage{natbib}
\usepackage{threeparttable}
\usepackage[titletoc,title]{appendix}
\usepackage{lipsum}
\usepackage{varwidth}
\usepackage{tabularx}
\newcommand{\etal}{\textit{et al}. }

\begin{document}
\title{Anisotropy-driven quantum capacitance in multi-layered black phosphorus}
\author{Parijat Sengupta}
\affiliation{Photonics Center, Boston University, Boston, MA 02215.}
\author{Shaloo Rakheja}
\affiliation{Dept of Electrical Engineering, New York University, Brooklyn, NY 11201.}

\begin{abstract}
We report analytic results on quantum capacitance (C$_{q}$) measurements and their optical tuning in dual-gated device with potassium-doped multi-layered black phosphorous (BP) as the channel material. The two-dimensional (2D) layered BP is highly anisotropic with a semi-Dirac dispersion marked by linear and quadratic contributions. The C$_{q}$ calculations mirror this asymmetric arrangement. A further increase to the asymmetry and consequently C$_{q}$ is predicted by photon-dressing the BP dispersion. To achieve this and tune C$_{q}$ in a field-effect transistor (FET), we suggest a configuration wherein a pair of electrostatic (top) and optical (back) gates clamp a BP channel. The back gate shines an optical pulse to rearrange the dispersion of the 2D BP. Analytic calculations are done with Floquet Hamiltonians in the \textit{off-resonant} regime. The value of such C$_{q}$ calculations, in addition, to its role in adjusting the current drive of an FET is discussed in context of metal-insulator and topological phase transitions and enhancements to the thermoelectric figure of merit.
\end{abstract}
\maketitle

\begin{figure}
\includegraphics[width=3.2in]{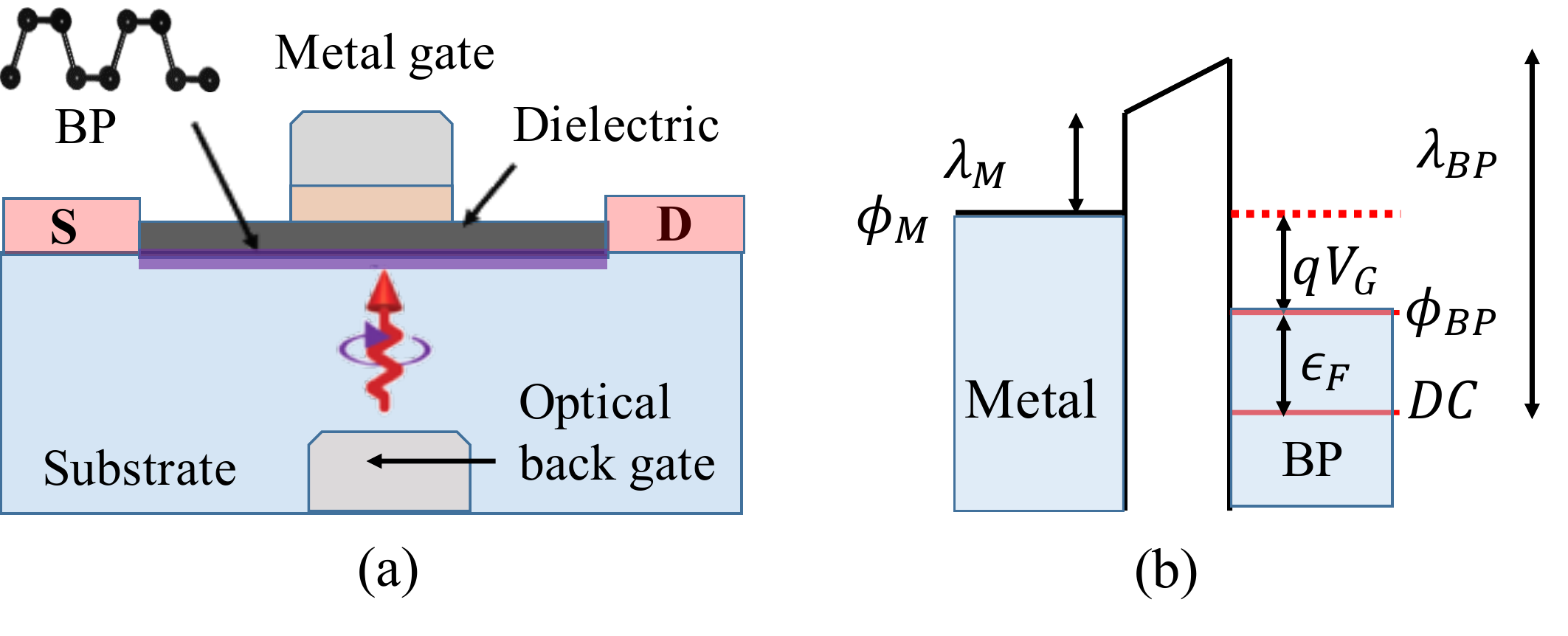}
\vspace{-0.3cm}
\caption{The left panel (a) shows the schematic of a metal-gated (V$_{g}$) four-layered \textit{K}-doped BP (with a puckered unit cell sketched in black) device with source (S) and drain (D) contacts. The energy edges are drawn in (b) where $ \varepsilon_{f}$ is the Fermi energy measured from the Dirac crossing. The electrochemical potential of the metal (BP) layer is $\phi_{M}\left(\phi_{BP}\right)$ while the work functions are identified by $\lambda$.}
\vspace{-0.55cm}
\label{schm}
\end{figure}

The quantum capacitance~\cite{luryi1988quantum} of a metal-oxide-semiconductor field-effect transistor (MOSFET) is usually masked by the classical oxide/dielectric capacitance and its contribution in evaluation of performance metrics is ignored. In a circuit representation, the quantum capacitance (C$_{q}$) 
appears in a 
series arrangement with the oxide capacitance (C$_{ox}$) and is discernible for values much lower than C$_{ox}$. The thinning of the oxide or introduction of a high-dielectric $\left(\kappa\right)$ insulating layer in state-of-the-art MOSFETs, therefore, suggests that the C$_{q}$ will influence~\cite{john2004quantum} the electrostatics and the overall capacitance; a demonstration of which has already been observed in measurements on Dirac materials.~\cite{yu2013interaction,xia2009measurement} As a case in point is the gated-graphene device which can be likened to a parallel-plate capacitor (see Fig.~\ref{schm} for a schematic) where the metal contact and the graphene sheet serve as the electrodes; the C$_{q}$ in such a prototypical arrangement has been reported to be a substantial fraction~\cite{ponomarenko2010density} of the total capacitance. The reduction in oxide thickness (or the use of a large $ \kappa$) aside, the C$_{q}$, which is essentially the response of the charge within the channel (in a MOSFET) as the conduction and valence band edges shift with a changing gate bias is significant for a linearly dispersing set of states, in contrast to a negligible contribution in case of a two-dimensional (2D) electron gas with a parabolic band description. A substantial C$_{q}$ has also been confirmed with other materials that host Dirac fermions characterized by a linear dispersion; for instance, Xiu \etal definitively observed Shubnikov-de Haas oscillations in C$_{q}$ measurements on Bi$_{2}$Se$_{3}$ thin films with topological helical surface states.~\cite{xiu2012quantum} 

In this letter, we examine the C$_{q}$ behaviour in multi-layered black phosphorous (BP), a relatively new addition to the family of 2D materials with a graphene-like honeycomb lattice~\cite{xia2014rediscovering} and thickness-dependent electronic structure. However, unlike graphene, multi-layer BP has a direct bulk band gap and marked by a distinct anisotropy evident in a puckered structure as it assumes an armchair and zigzag shape along \textit{x-} and \textit{y-} axes, respectively. From an applications perspective, experimental demonstrations of multi-layer BP MOSFETs with a large on/off-ratio, pronounced \textit{n}-channel transconductance~\cite{haratipour2016ambipolar}, and enhanced mobility~\cite{qiao2014high} exist. While other memebers of the 2D family exhibit similar behaviour, it is the intrinsic anisotropy of BP that permits a wider gamut of applications including a direction-dependent optical absorption, enhanced thermal sensitivity, and plasma oscillations. In particular, the resonant plasma frequency in BP shows a vectorial dependence~\cite{low2014plasmons} on momentum (graphene has a scalar relationship) allowing their tuning via optical polarization. Besides plasmons, another prominent illustration~\cite{kim2015observation} of anisotropy can be observed in the dispersion of a four-layer BP doped with potassium. The dispersion, which is Stark-effect mediated (arising from the internal electric field of \textit{K}-dopants), is quadratic and asymptotically linear along the zigzag and armchair directions, respectively. The hybrid of linear and quadratic bands, a semi-Dirac system~\cite{baik2015emergence} with a finite electric field tunable band gap and graphene-like point Fermi surface presents intriguing possibilities combining the innate anisotropy and benefits of a Dirac material. The anisotropy, beyond a certain doping threshold, manifests as a topological phase transition with a negative bulk band gap and protected edge states marked by tilted spin-polarized Dirac cones.~\cite{baik2015emergence} 

The C$_{q}$ measurements tied to the the density-of-states (DOS) is, therefore, a reasonable probe of the behaviour and any significant many-body phenomenon that governs the response of the electron ensemble in BP to external perturbations, such as an \textit{in-situ} electric field when \textit{K}-doped. The central theme of this letter is to examine an optical possibility of tweaking the distinctive anisotropy to explore avenues in improving device designs aided by a tunable charge or heat current. Quantum capacitance measurements exactly map the anisotropic DOS which mirrors the charge and heat response functions and are therefore useful in this regard. 
We begin our analysis by first defining $ C_{q} $ as $ e^{2}\mathcal{D}(\epsilon) $, the DOS in this case is limited to a region surrounding the anisotropic Dirac crossing (DC). The DOS, as we show through analytic calculations, is clearly a function of the inherent anisotropy; however, the anisotropy is further accentuated through irradiation with high-energy optical pulses such that photon-dressed band structure acquires a greater degree of asymmetry leading to a rise in the quantum capacitance. The rise is more pronounced for higher-energy levels. To accomplish this, we propose a dual-gated device structure, wherein the back gate (the top gate is the usual electrostatic-type) is equipped with a photon source whose energy and intensity levels can be continuously varied. Further applications are pointed out later including C$_{q}$ measurements as a marker for topological phase transitions in buckled honeycomb lattices and the optical tuning of DOS to enhance the thermoelectric figure of merit. As a clarifying note, while interface-trap capacitance (C$_{it}$) in a practical BP-based MOS setup is finite, it does not directly amend C$_{q}$; therefore, to a working approximation, we ignore such effects. 

The effective low-energy Hamiltonian for a four-layered black phosphorus sheet is~\cite{baik2015emergence}
\begin{equation}
\mathcal{H}_{p} = \left(\Delta + \alpha k_{x}^{2}\right)\sigma_{x} + \beta k_{y}\sigma_{y}.
\label{ham1}
\end{equation}
The band gap is $ 2\Delta $. The other terms are defined as $ \alpha = \hbar^{2}/2m^{*} $ and $ \beta = \hbar v_{f} $, where $ m^{*} $ is the effective mass for the parabolic branch and $ v_{f} $ is the Fermi velocity along the linear band. As the \textit{K}-doping increases, $ \Delta $ begins to shrink, eventually diminishing to zero. The Hamiltonian, $ \mathcal{H}_{d} $, is therefore just $ \alpha k_{x}^{2}\sigma_{x} + \beta k_{y}\sigma_{y} $. The subscripts \textit{p} and \textit{d} denote pristine and semi-Dirac \textit{K}-doped BP. Diagonalizing Eq.~\ref{ham1}, the eigen energy expressions in the Dirac semi-metal phase is
\begin{equation}
E\left(k\right) = \pm\,\sqrt{\left(\Delta + \alpha k_{x}^{2}\right)^{2} + \left(\beta k_{y}\right)^{2}}.
\label{bpdp}
\end{equation}
The +(-) sign in the energy expressions denote the conduction (valence) state. The dispersion of the doped sample by letting $ \Delta \rightarrow 0 $ in Eq.~\ref{bpdp} clearly points to massless Dirac Fermions along the armchair direction (\textit{y}-axis) while the zigzag axis (\textit{x}-axis) hosts its massive counterpart. For DOS calculation of anisotropic BP, we use the dispersion in Eq.~\ref{bpdp} to write $ \mathcal{D}(\epsilon) = \int \dfrac{d^{2}k}{\left(2\pi\right)^{2}}\delta\left(\epsilon - E(k)\right) = \dfrac{1}{\left(2\pi\right)^{2}}\sum\limits_{j}\int_{0}^{2\pi} d\theta \dfrac{k_{j}}{\vert g^\prime(k_{j})\vert} $. Here, the azimuthal angle is $ \theta $. We have used the identity $ \delta\left(g(x)\right) = \sum\limits_{j}\dfrac{\delta(x-x_{j})}{\vert g^\prime(x_{j})\vert}$ such that $ g(x_{j}) = 0 $ and $ x_{j} $ is a simple zero of $ g(x) $. The function $ g(k) $ in this $ \epsilon - \sqrt{\left(\Delta + \alpha k^{2}\cos^{2}\theta\right)^{2} + \left(\beta k\sin\theta\right)^{2}} $. Setting $ g(k) = 0 $ and solving, the positive root $\kappa $ can be expressed as $\kappa^{2} = \left(\sqrt{p^{2} + 4\alpha^{2}\cos^{4}\theta\left(\epsilon^{2} - \Delta^{2}\right)} - p\right)/\left( 2\alpha^{2}\cos^{4}\theta\right) $, where $ p = \left(2\Delta \alpha\cos^{2}\theta + \beta^{2}\sin^{2}\theta\right) $. Inserting the positive root and the derivative (w.r.t $ k $ at $ \kappa $) of $ g\left(k\right)$, the expression for DOS is
\begin{equation}
\mathcal{D}\left(\epsilon\right) = \dfrac{1}{4\pi^{2}}\int_{0}^{2\pi} d\theta\dfrac{E\left(k\right)}{2\alpha\cos^{2}\theta\left(\Delta + \alpha\kappa^{2}\cos^{2}\theta\right) + \left(\beta\sin\theta\right)^{2}}.
\label{dosfe} 
\end{equation}
The DOS expression in Eq.~\ref{dosfe} is evidently a function of $ \alpha $ and $ \beta $, the characteristic anisotropy markers of BP, and can be numerically evaluated. 

The two parameters reflecting the anisotropy can be modulated by an external perturbation, mostly mechanical strain or pressure; however, the analytic form of the dispersion (Eq.~\ref{bpdp}) remains unchanged. It would be therefore useful to check if there exists a possibility via an external control to alter the dispersion character. It is well-understood that interaction with a magnetic field (by forming Landau levels) rearranges the dispersion. The feasibility of such an approach in a circuit environment is modest as it must suppress any unintentional electromagnetic coupling. A workaround to achieveing change in the analytic description of the bands is suggested involving an optical back gate (in addition to the top metal gate in Fig.~\ref{schm}) fitted with a photon source that irradiates the BP channel. The light source is an external electromagnetic perturbation periodic in time. To quantitatively assess the change in dispersion under a periodic perturbation, we must make use of the Floquet theory~\cite{cayssol2013floquet} for a periodic Hamiltonian, $ \hat{H}\left(t\right) = \hat{H}\left(t + T\right) $ to construct the corresponding eigen states. Note that the time-dependent Hamiltonian is $ \hat{H}\left(t\right) = \hat{H}_{0} + \hat{V}\left(t\right)$, where $ \hat{H}_{0} $ is the stationary part and $ \hat{V}\left(t\right) $ is the time-periodic perturbation. The time-dependent Hamiltonian can be transformed into a Floquet Hamiltonian, which, following Tannor~\cite{tannor2007introduction}, is $ \hat{H}_{F}\left(t\right) =  \left(\hat{H}\left(t\right) - i\hbar\partial_{t} \right) $. The Floquet Hamiltonian when solved takes a matrix form~\cite{tannor2007introduction} and transforms a time-dependent problem to a time-independent Schr{\"o}dinger equation. For cases where the light frequency $ \left(\omega\right) $ is much higher than the energy scales of the stationary Hamiltonian, an approximation to the Floquet Hamiltonian can be written as~\cite{lopez2015photoinduced}
\begin{equation}
\hat{H}_{F}\left(k\right) = \hat{H}_{0} + \dfrac{\left[H_{-1}, H_{1}\right]}{\hbar\omega}.
\label{fleq1}
\end{equation}
The terms contained in the anti-commutator are the Fourier components which have the following generalized form: $ H_{m} = \dfrac{1}{T}\int_{0}^{T}dt \exp\left(im\omega t\right)H_{t} $. $ H_{t} $ is the time-dependent part of $ \hat{H}\left(t\right) $ and $ T = 2\pi/\omega $.

\noindent The immediate task, therefore, is to evaluate the Fourier components for the case of irradiated semi-Dirac BP, assuming the conditions of Eq.~\ref{fleq1} are true, and check for the DOS-reflected degree of anisotropy vis-\`a-vis Eq.~\ref{dosfe}. The Fourier components must be calculated using Eq.~\ref{ham1}. Changing into a time-dependent form by applying the Peierls substitution $ \hbar\mathbf{k} \rightarrow \hbar\mathbf{k} - e\mathbf{A}\left(t\right) $ that represents the coupling (via the vector potential $ \mathbf{A}\left(t\right) $) to the electromagnetic field, we rewrite Eq.~\ref{ham1} as 
\begin{equation}
\hat{H}\left(t\right) =  \left[\Delta + \dfrac{\alpha}{\hbar}\left(k_{x} + eA_{x}\left(t\right)\right)^{2}\right]\sigma_{x} + \dfrac{\beta}{\hbar}\left(k_{y} + eA_{y}\left(t\right)\right)\sigma_{y}.
\label{fleq3}
\end{equation}
The time-dependent part is therefore, $ \hat{H}_{t} = e\alpha/\hbar\left(2k_{x}A_{x}\left(t\right) + eA_{x}^{2}\left(t\right)\right)\sigma_{x} + \left(e\beta/\hbar\right) A_{y}\left(t\right)\sigma_{y} $. For circularly polarized light propagating along $ \hat{z} $, the two vector components are $ \mathbf{A} = A_{0}\left(-\sin\omega t\hat{x}, \cos\omega t\hat{y}\right) $. Here $ A_{0} = E_{0}/\omega $ and the electric field is $ \mathbf{E} = E_{0}\left(\cos\omega t\hat{x}, \sin\omega t\hat{y}\right) $. The vector potential and electric field are linked by the relation, $ \mathbf{E} = -\partial_{t}\mathbf{A} $. From the time-dependent part, the two Fourier components in Eq.~\ref{fleq1} $ \left(H_{\eta}, \eta = \pm 1 \right) $ can be evaluated. We have
\begin{equation}
\begin{aligned}
H_{\eta} &= \dfrac{1}{T}\int_{0}^{T}\biggl[\dfrac{\alpha\sigma_{x}}{\hbar}\left(-2A_{0}ek_{x}\sin\omega t + A_{0}^{2}e^{2}\sin^{2}\omega t\right) \\
& + \dfrac{e\beta A_{0}\sigma_{y}}{\hbar}\cos\omega t\biggr]exp\left(i\eta\omega t\right)dt, \\
&=  \dfrac{1}{T}\left[\dfrac{2i\eta \pi e\alpha A_{0}k_{x}\sigma_{x}}{\hbar\omega} + \dfrac{e\pi\beta A_{0}\sigma_{y}}{\hbar\omega}\right].
\label{comms}
\end{aligned}
\end{equation}
The commutator in Eq.~\ref{fleq1} using $ \left[\sigma_{x},\sigma_{y}\right] = 2i\sigma_{z} $ is
\begin{equation}
\mathcal{F} = \dfrac{\left[H_{-1}, H_{1}\right]}{\hbar\omega} = -\dfrac{1}{\hbar\omega}\dfrac{2\left(eA_{0}\right)^{2}k_{x}\alpha\beta}{\hbar^{2}}\sigma_{z}.
\label{commsf}
\end{equation}

\begin{figure}[t!]
\includegraphics[scale=0.55]{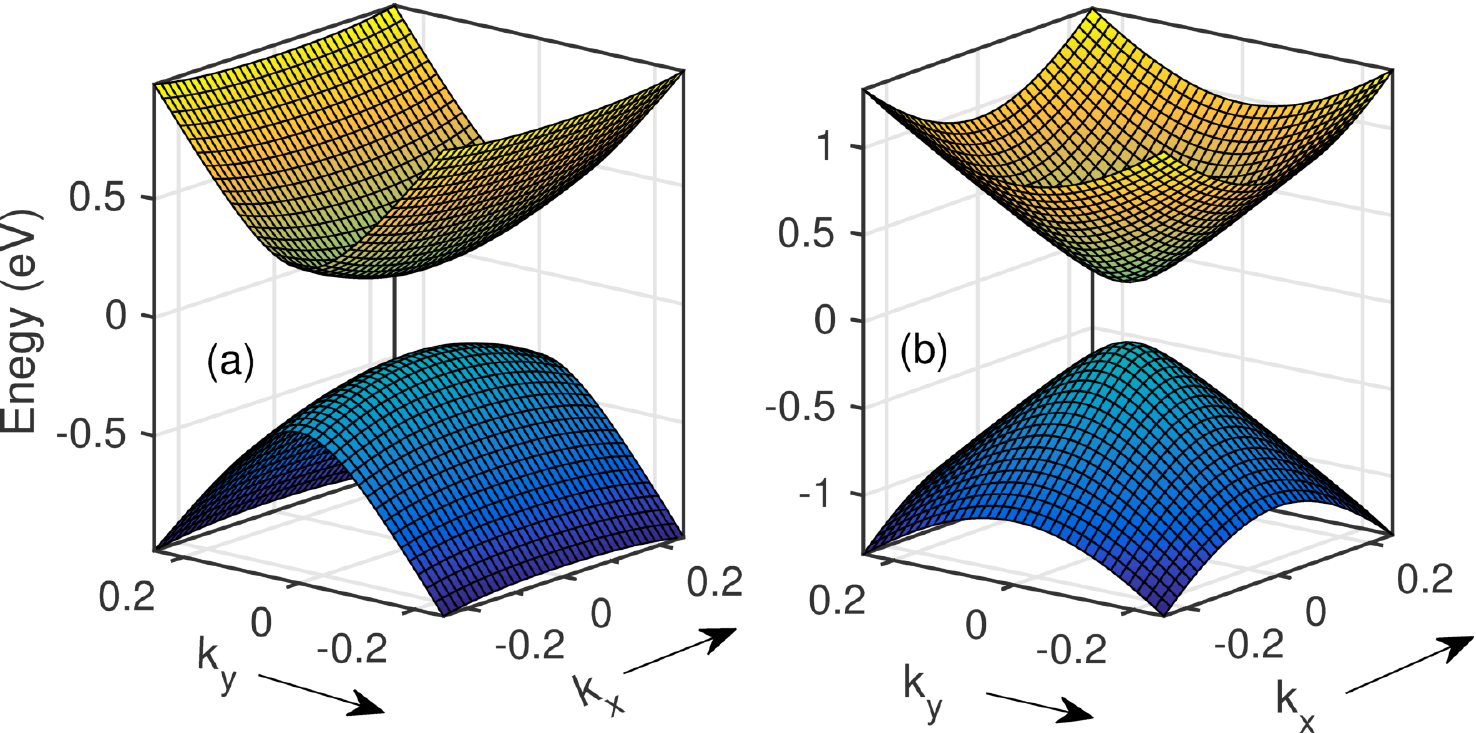}
\caption{The numerically calculated electronic dispersion of a four-layered BP in shown in the left panel (a) where the bands appear distinctly parabolic. However, under an intense light beam the bands are rearranged and acquires a linear character (b) as revealed in the funnel-like shape. For visual clarity, we have artificially set the intensity $\left( I \right) $ of the beam to a large value of $ 5.0\, eV $ while the frequency corresponds to an optical source of energy $ 7.0\, eV $. The $k$-components are in $\AA^{-1}$.} 
\vspace{-0.52cm}
\label{dpcmp}
\end{figure}

The rearranged Hamiltonian (superscript $ F $ indicates Floquet-modified) for four-layered BP using Eq.~\ref{commsf} is
\begin{equation}
H_{p}^{F} = \left(\Delta + \alpha k_{x}^{2}\right)\sigma_{x} + \beta k_{y}\sigma_{y} - \dfrac{1}{\hbar\omega}\dfrac{2\left(eA_{0}\right)^{2}k_{x}\alpha\beta}{\hbar^{2}}\sigma_{z}.
\label{flham}
\end{equation}
The dispersion by diagonalizing Eq.~\ref{flham} is
\begin{equation}
E^{F}\left(k\right) =  \pm\,\sqrt{\Omega^{2}k_{x}^{2} + \left(\Delta + \alpha k_{x}^{2}\right)^{2} + \left(\beta k_{y}\right)^{2}},
\label{fldp}
\end{equation}
where $ \Omega = 2\left(eA_{0}\right)^{2}\alpha\beta/\left(\hbar^{3}\omega\right) = 2\left(eA_{0}v_{f}\right)^{2}\left(\alpha/\beta\right)\left(\hbar\omega\right)^{-1} $. Before we proceed to evaluate the quantum capacitance from the DOS for the dispersion in Eq.~\ref{fldp}, it is useful to make several remarks; firstly, note that the additional Floquet-induced term is a function of the anisotropy ratio, $ \alpha/\beta $, and the original band gap $ \left(\Delta\right)$ at the zone centred $ \left( k = 0 \right) $ Dirac crossing remains preserved. The intrinsic anisotropy is disturbed with an additional linear $\left(k_{x}\right)$ term that is linked (through $ eA_{0}v_{f} $) to the energy of the irradiating beam and also imparts a greater linearity to the overall dispersion. For conditions, where the linear terms lend greater weight to the dispersion through Floquet inducements, there is a noticeable shift in character from a largely parabolic to a graphene-like cone structure. For a graphic description, see Fig.~\ref{dpcmp} and the accompanying caption. As for the DOS for a four-layer BP, which was inherently anisotropic to begin with, it now acquires a further measure of distortion for the photo-adjusted Floquet band dispersion and must manifest in quantum capacitance measurements. Finally, note that using the relation $ I = eA_{0}\omega/8\pi\alpha^{'}\hbar $, where $ I $ is the intensity and $ \alpha^{'} \approx 1/137 $ represents the fine structure constant, allows us to express the Floquet term in a particularly instructive form as : $\mathcal{F} = 8\pi\alpha^{'}Iv_{f}^{2}\left(\alpha/\beta\right)\omega^{-3} $. This suggests that $\mathcal{F} $ is intensity-adjustable (for a given frequency) and such control can be accomplished through an optical back gate, which reflects as a concomitant change in the observed dispersion and DOS.

\begin{figure}[t!]
\includegraphics[width=2.9in]{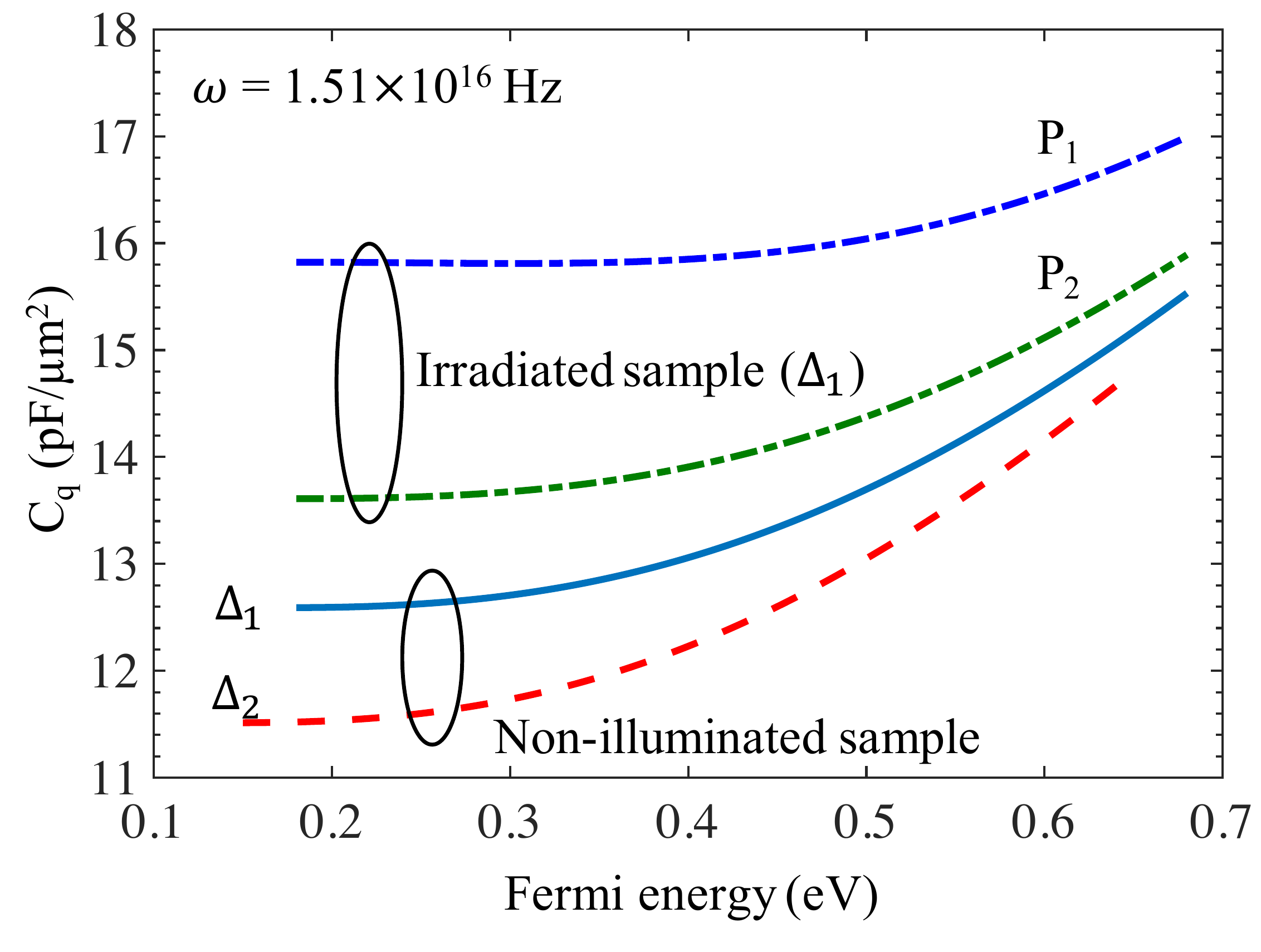} 
\vspace{-0.25cm}
\caption{The numerically computed C$_{q}$ in vicinity of the Dirac crossing $\left(0.18\,eV\right)$ is shown for non-illuminated and irradiated four-layered gapped BP (\textit{K}-doped) sample. Two sets of photo-illumination with different powers, $ P_{1} = 1.0 \, eV $ and $ P_{2} = 0.75 \, eV $, but identical frequency ($ \omega $) modulates the C$_{q}$. The non-illuminated C$_{q}$ is lower for a smaller band gap $\left(\Delta\right)$. Here, $ \Delta_{1} = 0. 18 \, eV $ and $ \Delta_{2} = 0. 15 \, eV $.} 
\vspace{-0.6cm}
\label{cqcomp}
\end{figure}

For C$_{q}$ calculations in the Floquet regime, the DOS from the dispersion in Eq.~\ref{fldp} can be written retracing the same steps as before for the pristine case. This gives $ \mathcal{D}^{F}\left(\epsilon\right) = \dfrac{1}{4\pi^{2}}\int_{0}^{2\pi}\gamma^{2}\dfrac{E^{F}\left(k\right)}{\Sigma}d\theta $, where $ \Sigma = \left\lbrace\left(\Omega^{2} + 2\Delta\alpha\right)\cos^{2}\theta + 2\gamma^{2}\alpha^{2}\cos^{4}\theta + \beta^{2}\sin^{2}\theta\right\rbrace $ and $ \gamma^{2} $ is a positive root of the bi-quadratic equation: $ \alpha^{2}k^{4}\cos^{4}\theta + \left(2\Delta\alpha\cos^{2}\theta + \beta^{2}\sin^{2}\theta + \Omega^{2}\cos^{2}\theta\right)k^{2} + \left(\Delta^{2} - \epsilon^{2}\right) = 0 $. We show in Fig.~\ref{cqcomp}, the numerically computed C$_{q}$ for two sets of circularly-polarized light, each identified by a specific power and frequency. A set of well-defined features are easily identifiable from the plot; first of all, C$_{q}$ is higher for an irradiated BP sample, the increase more noticeable for a beam with enhanced power at lower energies, suggesting that the Floquet term $\left(\mathcal{F}\right)$ dominates. A `dominating' Floquet term as evident from Fig.~\ref{dpcmp} bestows a greater degree of Dirac-character which reveals as an increment to the DOS and C$_{q}$. As we traverse in $k$-space away from the semi-Dirac crossing with an attendant rise in energy, the admixture of linear and quadratic terms undergoes a reformulation with the latter in the ascendancy allowing the weightier parabolic contribution to shrink the imparity between the curves in Fig.~\ref{cqcomp}. Essentially, it is an interplay between the intrinsic parabolicity and the combined linearity, in part due to the artificial Floquet-induced intensity-dependent Dirac-like term. We turn this interplay into a handle to adjust C$_{q}$ via optical power. Further, from Eq.~\ref{fldp}, since the derived Floquet term is anisotropy ratio $\left(\nu\right)$ regulated, quantum-confinement, that has been demonstrated before~\cite{2015quantum} in adjusting $ \nu $, in conjunction, with intensity-levels can augment or reduce C$_{q}$. In addition, in experimental measurements, the minimum of the C$_{q}$ curve points to the Dirac crossing and locates its exact position. lastly, the radiation-induced band rearrangement and an enhanced DOS can be a comparative optical analog to chemical functionalization that introduces `controlled' energy levels in material systems.  

While alterations to C$_{q}$ is useful from the standpoint of controlling the current drive of a FET, ancillary but pertinent data such as the surface potential can be extracted. Traditionally, low-frequency capacitance measurements allow approximating the surface state potential by a combination of an ac signal superimposed on $ V_{a} $, a dc bias . A simple analytic formulation of this reads~\cite{berglund1966surface}
\begin{equation}
\Psi_{s}\left(V_{2}\right) - \Psi_{s}\left(V_{1}\right) = \int_{V_{1}}^{V_{2}}\left[1 - \dfrac{C\left(V_{a}\right)}{C_{ox}}\right]dV_{a}.
\label{bgi}
\end{equation}
In Eq.~\ref{bgi}, $ C\left(V_{a}\right) $ and $ C_{ox} $ are the total measured and oxide/dielectric capacitance, respectively while the gate bias is swept between $ V_{1} $ and $ V_{2} $. A similar form of the integral in Eq.~\ref{bgi} in context of optical fields can be written; recall that holding the frequency constant and varying the power level through the back gate, the Floquet term changes C$_{q}$ (see Fig.~\ref{cqcomp}) and is mirrored in the final capacitance data through the relation $ C = C_{q}^{-1} + C_{ox}^{-1} $. The integral in Eq.~\ref{bgi} can therefore be modified such that original quantities are now a function of the power of photo-illumination; for instance, $ \Psi_{s}\left(V\right) \rightarrow \Psi_{s}\left(P\right) $. Likewise, this idea could also be potentially extended by replacing the back-gated light source, in a crude way, by a radio-frequency (RF) generator to assess the RF-governed C$_q$ of a Dirac material.

While the focus heretofore has been on the utility and tunability of C$_{q}$ as a circuit element, the footprint can be expanded. A prime example is in the area of capacitance spectroscopy data which can reveal vital information such as a metal-insulator transition in case of MoS$_{2}$.~\cite{chen2014probing} As we briefly alluded to in the opening section, \textit{K}-doped BP can be electric field driven to phase transition to a topological insulator (TI); similar situations also exist for graphene-analogs, for example, in the 2D material, silicene, the electric field controls the opening and closing of band gap as the phase toggles between a trivial band insulator, valley-polarized metal, and TI.~\cite{ezawa2013photoinduced} Each such transition can be reflected in their corresponding C$_{q}$ measurements that conceivably require a lesser degree of experimental sophistication (and impervious to scattering details) in contrast to higly precise experiments in the topological transport regime to observe helical Dirac fermions.~\cite{hsieh2009tunable} Moreoever, the tunable anisotropy of BP can be gainfully employed in designing thermal devices~\cite{qin2014hinge} with a significant thermoelectric figure of merit (\textit{ZT}) which is large for high-electric and low-thermal conductivities. Application-specific anisotropic electric and thermal  conductances can be achieved through simple modulations to the DOS through photo-illumination.

\bibliographystyle{apsrev}

\end{document}